\documentstyle[preprint,aps,amsfonts]{revtex}
\def\vec#1{{\bf#1}}
\def\hatn#1{\hat{\bf#1}}
\def\mat#1{\stackrel{{ }_{\,\leftrightarrow}}{#1}}
\begin{document}
\draft
\title{Explanation for the Resistivity Law  in Quantum Hall Systems}
\author{Steven H. Simon and Bertrand I. Halperin}
\address{Department of Physics, Harvard University, Cambridge, MA
  02138}

\date{Submitted June 6 1994, Revised August 8 1994}

\maketitle

\begin{abstract}
  \noindent We consider a 2D electron system in a strong magnetic
  field, where the local Hall resistivity $\rho_{xy}(\vec r)$ is a
  function of position and $\rho_{xx}(\vec r)$ is small compared to
  $\rho_{xy}$.  Particularly if the correlations fall off slowly with
  distance, or if fluctuations exist on several length scales, one
  finds that the macroscopic longitudinal resistivity $R_{xx}$ is only
  weakly dependent on $\rho_{xx}$ and is approximately proportional to
  the magnitude of fluctuations in $\rho_{xy}$.  This may provide an
  explanation of the empirical law $R_{xx} \propto B
  \frac{dR_{xy}}{dB}$ where $R_{xy}$ is the Hall resistance, and $B$
  is the magnetic field.
\end{abstract}



\narrowtext

Almost ten years ago, it was noticed that the longitudinal resistance
$R_{xx}$ of a quantum Hall system looks very much like the
derivative of the Hall resistance $R_{xy}$ with respect to filling
fraction\cite{Chang}.  More recent experiments\cite{Stormer} have
shown that in high mobility samples, the relation
\begin{equation}
  \label{eq:chang}
  R_{xx} = \alpha B \frac{dR_{xy}}{dB}
\end{equation}
with $B$ the magnetic field, and $\alpha$ a constant, holds amazingly
well over a wide range of temperatures and magnetic fields, including
both the integer and fractional quantized regimes, and unquantized
regions where $R_{xy}$ varies linearly with $B$, as for a classical
system.  The constant $\alpha$ is sample dependent, and varies
somewhat with temperature, but is typically in the range $10^{-1}$ to
$10^{-2}$.  Despite the simplicity and generality of this empirical
relation, it has defied explanation for almost a decade.  In this
paper, we propose an explanation based on inhomogeneities in density
over long length scales.

We consider a two dimensional electron system where the local electron
density $n(\vec r)$ is fixed by the charged impurity distribution in a
nearby doping layer\cite{Efros}.  We assume that the fluctuations in
the density $\delta n(\vec r) = n(\vec r) - n_0$ are much smaller than
the average density $n_0$, and we define a dimensionless correlation
function $g$ such that $\langle \delta n(\vec r) \delta n(\vec r')
\rangle = \mu^2 g(|\vec r - \vec r'|)$ with $g(0) = 1$, and $g(x)
\rightarrow 0$ as $x \rightarrow \infty$.  Although we will later be
concerned with the effects of disorder over several length scales
simultaneously, to begin with, we will consider the case where there
is only disorder on a single length scale $a$.  As an example of a
correlation function with a single length scale, we can consider $g(x)
= \exp(-[\frac{x}{a}]^2)$.  Typically, the length scale $a$ of any
disorder will be greater than or equal to the setback distance of the
doping layer.  If $a$ is sufficiently long, we can define a {\it
  local} resistivity tensor $\mat \rho(\vec r)$ which depends on the
local conditions.  We assume that at a fixed temperature $T$, the
local Hall resistivity $\rho_{xy}(\vec r) = - \rho_{yx}(\vec r)$ is
purely a function of the local filling fraction $\nu(\vec r) \equiv
\phi_0 n(\vec r)/B$ with $\phi_0 = hc/e$.  This assumption is
certainly valid at high temperatures, where $\rho_{xy}$ takes on the
classical value, $\rho_{xy}(\vec r) = e^2 \nu(\vec r)/h$.  The
assumption should also be valid at least as a first approximation in
the ranges of $T$ and $B$ where the integer or fractional quantized
Hall plateaus begin to develop.  Then, if the density fluctuations are
not too large, we may write $\rho_{xy}(\vec r)= f(\nu(\vec r)) =
\rho_0 + \delta \rho_{xy}(\vec r)$ where $\rho_0 = f(\nu_0)$, $\nu_0 =
\phi_0 n_0/B$, and $\delta \rho_{xy}(\vec r) = f'(\nu_0) \phi_0 \delta
n(\vec r)/B$.  Thus we have $\langle \delta \rho_{xy} \rangle = 0$ and
$\langle \delta \rho_{xy} (\vec r) \delta \rho_{xy} (\vec r') \rangle
= \lambda^2 g(|\vec r - \vec r'|)$, with $\lambda = f'(\nu_0) \phi_0
\mu /B$.  We also assume that there is a local diagonal resistivity
$\rho_{xx}(\vec r)$ which may depend on the temperature and the
magnetic field, but which is small compared to $\lambda$.  This
assumption should be quite reasonable for high mobility samples.  For
simplicity, we shall take $\rho_{xx}(\vec r) = \tilde \rho$,
independent of $\vec r$.

We now want to find the macroscopic resistivity $R_{xx}$ for the
system.  The current profile $\vec j(\vec r)$ is
defined by Maxwell's equation $\nabla \times \vec E = 0$ and current
conservation.  Using $\vec E = \mat \rho \vec j$, we can thus define
the current with the system of equations
\begin{equation}
\label{eq:sys}
\vec \nabla \times (\mat
\rho \vec j) = 0$ ~~~ \, , ~~~ $\vec \nabla \cdot \vec j = 0,
\end{equation}
along with appropriate boundary conditions and the condition that the
spatial average of $\vec j$ have a specified value $\vec j_0$.  By
using the second equation of (\ref{eq:sys}), we can rewrite the first
equation as
\begin{equation}
\label{eq:sys2}
(\vec \nabla \rho_{xy} \times \vec j)
+ \tilde \rho (\vec \nabla \times \vec j) = 0,
\end{equation}
which is clearly independent of $\rho_0$.  The total dissipation per
unit area is given by $\frac{1}{2} R_{xx} |\vec j_0|^2$.  Since the
local dissipation $\frac{1}{2} \tilde \rho |\vec j|^2$ is independent
of $\rho_0$, $R_{xx}$ must also be independent of $\rho_0$.  On
dimensional grounds (since $R_{xx}$ depends only on $\lambda$, $\tilde
\rho$, and the function $g$), we expect that in the limit $\tilde \rho
\rightarrow 0$, the longitudinal resistance should have a scaling form
\begin{equation}
R_{xx} = C \lambda^{1 - \omega} \tilde \rho^\omega
\end{equation}
where $C$ is a nonuniversal constant that depends on the form of the
function $g$.  Calculations described below suggest that the exponent
$\omega$ can be close to zero, particularly if we consider the
effects of disorder on several length scales simultaneously.  If
indeed we find that $\omega \approx 0$, then we have $R_{xx} \approx C
\lambda = C f'(\nu_0) \phi_0 \mu /B$.  On the other hand, the Hall
resistance will be given by $R_{xy} = \rho_0 + {\cal O}(\lambda^2)
\approx f(\nu_0)$ so $B \frac{d R_{xy}}{dB} = f'(\nu_0) \phi_0 n_0/B$,
thus establishing Eq.  (\ref{eq:chang}) with $\alpha = C \mu/n_0$.

A different model, which may be more applicable for high mobility
systems at temperatures that are not too low, has dissipation arising
from electron-electron scattering, which causes no dissipation for a
uniform flow velocity.  If the magnetic field is not too strong, we
may represent this by an electron fluid viscosity $\eta$, instead of a
resistivity $\tilde \rho$, so that the local dissipation is given by
$\eta n^2|\vec \nabla \times (\vec j/n)|^2$ rather than $\tilde \rho |
\vec j|^2$, with $\vec j$ the local current density.  In this case we
find the similar result $R_{xx} = C a^{-2\omega} \lambda^{1 - \omega}
\eta ^\omega$.  If the exponent $\omega$ is sufficiently close to
zero, so that we may neglect the field-dependence of $\eta$, we may
again derive the resistivity law (\ref{eq:chang}).

The conductivity problem described above has been studied by various
authors\cite{IsiReview,Shklovskii,Fogler}.  These analyses conclude
that the exponent $\omega$ for the case of a local resistivity $\tilde
\rho$, with finite range correlations in $\rho_{xy}$, is $\omega =
\frac{3}{13} \approx 0.23$.  A similar analysis for the case of a
local viscosity gives $\omega= \frac{3}{19}$, as will be discussed
below.  (This result has also been derived by Fogler and
Shklovskii\cite{Fogler}.)

A starting point for understanding this conductivity problem is the
observation that in the limit $\tilde \rho \rightarrow 0$, the current
$\vec j(\vec r)$ is confined to a contour of constant $\rho_{xy}(\vec
r)$ which percolates across the system\cite{IsiReview,Shklovskii}.  If
$\delta \rho_{xy}(\vec r)$ is statistically symmetric about zero, then this
percolating
contour\cite{Trugman} is defined by $\rho_{xy}(\vec r) = \rho_0$.  As
$\tilde \rho$ is increased, the current spreads out somewhat from this
contour.  The physical interpretation of $\omega \approx 0$ is that
(in either the viscous case or the resistive case) as we decrease the
dissipative part of the local resistivity (fixing the potential
difference across the system), the current profile readjusts such that
the total dissipation remains approximately fixed.

If we imagine one of the percolating contours of $\rho_{xy}(\vec r) =
\rho_0$ to be locally pointing in the $\hatn y$ direction, we can
write locally $\delta \rho_{xy}(x,y) = x Q(y)$ for some function $Q$.  Since we
suspect that the current will follow along this contour, we propose
that the $\hatn y$ component of the current can be written in the form
$\vec j_y(x,y) = J S[x/w(y)]/w(y)$ where $J$ is the total current
carried by the channel, $w(y)$ is the local width of channel, and $S$
is an unknown function satisfying the normalization condition $\int dz
S[z] =1$.  Of course, once we know $\vec j_y$, $\vec j_x$ is
determined by current conservation.  In the small $w$ limit it is easy
to show that a solution to the system of equations (\ref{eq:sys}) is
given for the functional form $S[z] = \exp(-z^2/2)/\sqrt{2\pi}$ when
the width $w$ of the channel is determined by the differential
equation
\begin{equation}
\label{eq:diff}
d [(w Q)^p]/dy = - p \, \tilde \rho \, Q^{p-1}
\end{equation}
where $p=2$ here.  In the case of viscous dissipation, we find a
differential equation similar to (\ref{eq:diff}), except that $\tilde
\rho$ is replaced by $\eta$, and $p=4$.  In the viscous case, the
function $S$ must be chosen to satisfy the differential equation
$d^3S(z)/dz^3 = -z S(z)$.

As an explicit example, we consider the periodic geometry $\delta
\rho_{xy}(\vec r) = \lambda \left[\cos\left(\frac{\pi (x+y)}{s}\right)
-\cos\left(\frac{\pi (x-y)}{s}\right) \right]$ which has a square
lattice of saddle points at $(\frac{x}{s},\frac{y}{s}) \in {\Bbb
  Z}^2$, and we take $\tilde \rho \ll \lambda \ll \rho_0$.  We assume
an average current $\vec j_0$ in the $\hat y$ direction so that the
current flows down channels of fixed $\rho_{xy}(\vec r) = \rho_0$
along the $x = \mbox{integer}$ lines.  Near the channel at $x=0$, for
example, we write $\delta \rho_{xy}(x,y)=x Q(y)$ with $Q(y) = - H
\sin(\pi y/s)$, where $H$ is the characteristic perpendicular slope of
the function $\delta \rho_{xy}$.  In this geometry we have $H = 2 \pi
\lambda/s$, but in a more general geometry, $H \sim \lambda/a$, where
$a$ is the characteristic length scale of the density fluctuation.

Integrating $w$ from one saddle to the next in Eq. (\ref{eq:diff}), we
find that the width of the channel is given by $w(y) = \sqrt{\tilde
  \rho s/\pi H} |\sec (\pi y/2 s)|$ in the resistive case.  Of course
the small $w$ approximation breaks down near the points where this
expression diverges, so the solution must take a different form very
close to these points.  The important thing to extract from this
result is that the characteristic width $W$ of the channel scales as
$W \sim \sqrt{\tilde \rho s/H}$.  Numerical
solutions\cite{Unpublished1} of the system of equations (\ref{eq:sys})
to find the exact current profile in this periodic geometry (as well
as in other simple geometries) support this conclusion.  Since we have
determined the current profile, we can also calculate the local
dissipation $\tilde \rho |\vec j|^2$ and thus extract the resistance
of this current carrying channel. In general, we find that the
resistance of a channel of length $s$ scales as $R_s \sim \tilde \rho
s/ W \sim H W$.  A similar argument for the case of viscous
dissipation yields the results $W \sim (\eta s/H)^{\frac{1}{4}}$ and
$R_s \sim \eta s/ W^3 \sim H W$.

We must remember, however, that we do not actually have a regular
array of saddle points.  We have instead a tortuous conduction
network.  We define a percolation problem by choosing a cutoff
$\epsilon$, considering all points such that $|\delta \rho_{xy}(\vec
r)|/\lambda < \epsilon$ to be ``conducting'' points, and all other
points to be ``insulating.''  This type of problem\cite{Trugman} has a
critical value of $\epsilon_c = 0$, provided we take $\delta
\rho_{xy}$ to be statistically symmetric about zero.  Our actual
system will have thin channels of width $W$ centered at the
percolating $\delta \rho_{xy}=0$ lines, and we will require for self
consistency that $\epsilon \sim W/a$ for these channels.  The
percolation network has a web-like topology consisting of essentially
one dimensional conduction paths that snake through the system and
intersect each other once every distance $\xi$.  More properly, the
correlation length $\xi$, is defined as the size of the largest
insulating islands in the conducting network.  As we approach the
percolation threshold by letting $\epsilon \rightarrow 0$, the
conduction channels become thinner, connections between conduction
paths are broken, and the correlation length diverges as $\xi \sim a
\epsilon^{-\nu}$.  For the case where the disorder has only one length
scale, it is known that $\nu = \frac{4}{3}$
\cite{IsiReview,Trugman,Stauffer}.

Although the conduction path in this percolation problem can branch
somewhat\cite{IsiReview,Trugman,Stauffer} on scales smaller than $\xi$,
forming dead-end loops and small multiply connected regions, it is
assumed for now that the current effectively follows a one dimensional
path around the perimeter of these branched region
\cite{IsiReview,Shklovskii}.  The length of such a path connecting
points separated by a distance $\xi$ scales as $a(\xi/a)^D$ where $D$
is the fractal dimension of the path.  For this particular problem,
where there is only one length scale, it is believed that
$D=\frac{7}{4}$\cite{IsiReview}.

Due to the web-like topology of the network, we can think of the
percolation network as an effectively periodic system similar to the
one we analyzed above\cite{IsiReview,Shklovskii}.  Here the
periodicity length is $\xi$.  However, since the channels are not
straight, the length $s$ of the conduction channels must be taken to
be the typical path length $a(\xi/a)^D$ between intersections of
macroscopic conduction paths.  Thus we have $W \sim \sqrt{\tilde \rho
  s/H} \sim a \sqrt{\tilde \rho \xi^D/\lambda a^D}$.  On the other
hand, our self consistency condition requires $W \sim a \epsilon$.
Using $\xi \sim a \epsilon^{-\nu}$, we find $\tilde \rho
\sim \lambda \epsilon^{2 + \nu D}$.  Since the system is homogeneous
on length scales larger than $\xi$, and since there is typically only
one conduction path per square of length $\xi$ on a side, the
macroscopic longitudinal resistivity is given by $R_{xx} \sim R_s \sim
HW \sim \lambda \epsilon \sim \lambda^{1- \omega} \tilde \rho^\omega$
where $\omega = \frac{1}{2 + \nu D} = \frac{3}{13}$ in the case of
a single disorder length scale.  This result agrees with the
previous work of Refs.~\cite{IsiReview} and \cite{Shklovskii}.  An analogous
calculation can be performed for the viscous case with a single
disorder length scale (using $W \sim [\eta s/H]^{\frac{1}{4}}$ instead
of $W \sim \sqrt{\tilde \rho s/H}$) to give the result $\omega =
\frac{1}{4+\nu D} = \frac{3}{19}$, in agreement with
Ref.~\cite{Fogler}.

Although these arguments seem quite reasonable, and are in decent
agreement with existing numerical work\cite{IsiReview}, it is possible
that they are not exact.  In particular, we are not very certain of
the fractal dimension $D$ of the conduction path.  For example, if the
current were to readjusts itself only very slightly so as to cut off
dead-end loops, $D$ might be changed.  However, it is clear that $1
\le D \le 2 $, and this uncertainty allows only a small range of
values of $\omega$ ($\frac{3}{14} \le \omega \le \frac{3}{10}$ in the
resistive case, and $\frac{3}{20} \le \omega \le \frac{3}{16}$ in the
viscous case).

We now consider the effect of having disorder on multiple length
scales.  As an extreme example, let us consider a system that is
disordered on a microscopic length scale $a$ (as above) and also
disordered on a macroscopic length scale $a'$ such that $a \ll a' \ll
L$, where $L$ is the size of the system.  On scales much less than
$a'$, the calculation described above should hold.  I.e., given a
local resistivity tensor $\mat \rho$ such that $\rho_{xy}$ has some
typical fluctuation $\lambda$, and $\rho_{xx} =\tilde \rho$, we can
use the above approach to calculate that the resistivity tensor $\mat
\rho'$ measured on a scale larger than $\xi$ but smaller than $a'$ is
given by $\rho'_{xx} = \tilde \rho ' \sim \lambda^\frac{10}{13} \tilde
\rho^\frac{3}{13}$ with $\rho'_{xy}$ given by the local mean value of
$\rho_{xy}$.  We can then repeat the argument to account for disorder
on the scale $a'$ by using the tensor $\mat \rho'$ as a local
resistivity.  Here, we similarly find that the macroscopic resistivity
is given by $R_{xx} \sim \lambda^\frac{10}{13} (\tilde
\rho')^\frac{3}{13} \sim \lambda^\frac{160}{169} \tilde \rho^\frac{9}{169}$.
Thus the exponent $\omega$ is effectively squared if disorder exists
on two very different length scales.  Clearly, this approach can be
extended analogously to account for disorder on any number of well
separated length scales, with the result that the exponent $\omega$
can become arbitrarily small if the disorder extends over a
sufficiently wide range of scales -- i.e., if the correlations fall
off sufficiently slowly.  Isichenko has previously considered the effect
of long range correlated disorder and has similarly found that $\omega$ can be
arbitrarily small for certain correlations that fall off very
slowly\cite{IsiReview}.

We can also show rigorously for a slightly modified model, that the
exponent $\omega$ can never be negative.  In the modified model, we
take $\rho_{xy}(\vec r) = \rho_0 e^{h(\vec r)}$, and $\rho_{xx}(\vec
r) = \tilde \rho e^{h(\vec r)}$, where $h(\vec r)$ is dimensionless
here and has a symmetric distribution about zero.  This model is
statistically self dual, in the sense that for any member $A$ of the
ensemble with a local resistivity tensor $\mat \rho_A\!(\vec r)$,
there is another $B$, having equal probability, with $[\mat
\rho_B\!(\vec r)]^{\dagger} = [\mat \rho_{A}\!(\vec r)]^{-1} (\rho_0^2
+ \tilde \rho^2)$.  It can be shown that the ensemble-averaged
resistivity tensor $\mat R$ satisfies $\mat R^{\dagger} = \mat R^{-1}
[\rho_0^2 + \tilde \rho^2]$, which implies $R_{xx}^2 + R_{xy}^2 =
(\rho_0^2 + \tilde \rho^2)$. (The proof employs a transformation
analogous to that used by Dykhne and Ruzin\cite{Dykhne}, with the
choice $a=c=0$, $d=b^{-1}=[\rho^2 + \tilde \rho^2]^{1/2}$ in Eq. [8]
of Ref. \cite{Dykhne}.)  Therefore, $R_{xx}$ cannot diverge for $\tilde
\rho \rightarrow 0$, and we must have $\omega \ge 0$.  More generally,
one expects that the exponent $\omega$ will always be $\ge 0$ for any
model where the macroscopic resistivity is isotropic\cite{IsiReview}.

The hypothesis that disorder exists on several length scales seems
plausible.  Although one certainly expects to find disorder on the
scale of the dopant setback distance, there is also evidence for
inhomogeneities in the electron density on the scale of a few
percent\cite{WillettPRL93}, and the length scale of these
inhomogeneities may be relatively large.  In this regard, it is
interesting that the values of $\sigma_{xx}$ obtained from the surface
acoustic wave experiments of Willett et.~al.~using a conventional
theoretical analysis, in the low frequency limit, are systematically
smaller by a factor of $\approx 3$ than the values obtained from
macroscopic conductivity measurements on similar samples, over a range
of fields and temperatures.  (This discrepancy was removed by an
explicit renormalization of a parameter $\sigma_m$ in the
experimental analysis\cite{WillettPRL93,HLR}).  If the sample has
sufficiently large density fluctuations on a length scale large
compared to the ultrasound wavelength together with smaller
fluctuations on a shorter length scale, this could conceivably account
for the observed discrepancy.  However, there remains a problem in
accounting for the experimental values of the coefficient $\alpha$ in
Eq.  (\ref{eq:chang}) via large scale inhomogeneities.  Analysis by
Dykhne and Ruzin\cite{Dykhne} in the case where $\rho_{xy}(\vec r)$
can only take on two discrete values, $\rho_1$ and $\rho_2$, gives a
maximum value of $R_{xx}$ equal to $|\rho_1 - \rho_2|/2$, when the
value of $\tilde \rho$ is small.  Assuming this estimate to be an
upper bound on $R_{xx}$ for the continuum case, we see that in order
to obtain a coefficient $\alpha \approx 0.03$, as in typical
experiments\cite{Stormer}, one would need to have large scale density
fluctuations of order $6$ percent, which is probably unreasonably
large.

If the exponent $\omega$ were precisely zero, then the resistivity law
(\ref{eq:chang}) would hold independent of the microscopic details of
the local dissipative resistivity (or viscosity) so long as the local
dissipation is sufficiently small.  If $\omega$ is small, but nonzero,
then the accuracy of the law will be determined to a some extent by
the behavior of the local resistivity.  In the high temperature
regime, for example, if the local resistivity $\tilde \rho$ itself
varies as $B^\alpha$ with $\alpha \approx \frac{1}{2}$, then at high
temperatures the macroscopic resistivity would vary as $B^{1 -
  \omega/2}$.

Theoretical results that in various particular circumstances, within
the quantum Hall regime, the macroscopic longitudinal resistivity is
determined by local fluctuations of the electron density $n(\vec r)$,
or equivalently of the local Hall resistance $\rho_{xy}(\vec r)$, have
been discussed previously by a number of
authors\cite{Shklovskii,Fogler,Dykhne,HLR,Others}.  Much of this
previous work has concentrated on the narrow transition regions
between neighboring well-developed Hall plateaus at low temperatures.
In the present note, we have been more concerned with the regime of
high temperature, where $\rho_{xy}(\vec r)$ may be more properly
considered to be a continuous function of the local filling factor and
hence a continuous function of position in the sample.  Moreover, we
have shown that a percolation analysis of the conductivity problem may
be used to explain the empirical resistivity law (\ref{eq:chang})
which should hold particularly well if disorder occurs over a wide
range of length scales.  Further analysis of the inhomogeneities
occuring in actual samples is clearly necessary to complete this
argument, as is a better understanding of dissipation processes at the
shortest length scales.

The authors are very grateful to B.~I.~Shklovskii for criticisms of an
earlier version of this paper, and for calling our attention to
several key references.  Enlightening discussions with H.~Stormer,
R.~L.~Willett, P.~A.~Lee, D.~B.~Chklovskii, C.~L.~Henley, and
N.~R.~Cooper are also acknowledged.  This work was supported by the
National Science Foundation Grant DMR-91-15491.


\end{document}